\documentclass[a4,twocolumn,showpacs,preprintnumbers,amsmath,amssymb]{revtex4}

\usepackage{graphicx}

\begin{document}
\title{\narrower Bistability and resonance in the periodically stimulated\\
Hodgkin-Huxley model with noise}

\author{L.~S.~Borkowski}
\affiliation{Department of Physics, Adam Mickiewicz University,
Umultowska 85, 61-614 Poznan, Poland}

\begin{abstract}
We describe general characteristics of the Hodgkin-Huxley
neuron's response to a periodic train of short current pulses
with Gaussian noise.
The deterministic neuron is bistable for antiresonant frequencies.
When the stimuli arrive at the resonant frequency
the firing rate is a continuous function of the current amplitude $I_0$
and scales as $(I_0-I_{th})^{1/2}$, characteristic of a saddle-node bifurcation
at the threshold $I_{th}$.
Intervals of continuous irregular response alternate
with integer mode-locked regions with bistable
excitation edge.
There is an even-all multimodal transition between
the 2:1 and 3:1 states in the vicinity of the main
resonance, which is analogous to the odd-all transition
discovered earlier in the high-frequency regime.
For $I_0<I_{th}$ and small noise the firing rate
has a maximum at the resonant frequency.
For larger noise and subthreshold stimulation
the maximum firing rate initially shifts toward lower frequencies,
then returns to higher frequencies in the limit of large noise.
The stochastic coherence antiresonance, defined as
a simultaneous occurrence of (i) the maximum
of the coefficient of variation, and (ii) the minimum
of the firing rate vs. the noise intensity,
occurs over a wide range of parameter values,
including monostable regions.
Results of this work can be verified experimentally.
\end{abstract}
\maketitle

\section{Introduction}
The Hodgkin-Huxley (HH) model \cite{HH1952}
is a prime example of a resonant neuron.
It was originally developed to explain experimental
properties of the squid giant axon.
Its behavior under the influence of constant,
periodic \cite{Holden1976,Guttman1980,Read1996,Chik2001,Lee-kim2006}
and irregular external current \cite{Hasegawa2000}
has been studied extensively.
It also served as a starting point for a number
of reduced models \cite{FitzHugh1961,Nagumo1962,Izhikevich2006},
designed to preserve the key features, while being more
amenable to large-scale numerical simulations.
However, a full understanding
of all qualitative properties of its solutions
has not been achieved yet.

Deterministic HH equations can have both periodic
and aperiodic, sometimes chaotic solutions \cite{Guckenheimer2002}.
Theoretical \cite{Rinzel1981,Winfree1987,Cronin1987} and
experimental \cite{Guttman1980,Winfree1987} analysis revealed
that near the excitation threshold two solutions,
the fixed point and the limit cycle, may coexist.
A simple such example is the HH model driven by a constant current.
As the current magnitude is increased the neuron
starts responding at a preferred frequency, which
is close to 50 Hz for the original HH parameter set.

The HH neuron in the presence of noise may display either resonant
or antiresonant behavior, depending on the signal magnitude, frequency, and noise.
The enhancement of weak signals by noise is known
as stochastic resonance \cite{Gang1993,Pikovsky1997,Longtin1997}.
The opposite effect, stochastic antiresonance, 
in which a neuron's firing frequency is slowed down
or even entirely stopped at some intermediate noise level,
received some attention \cite{Gutkin2008,Gutkin2009,Borkowski2010}
recently. 
Depending on model parameters this behavior
is associated with bistability \cite{Gutkin2008,Gutkin2009}
or multimodality of the response \cite{Borkowski2010b}.
A relation between bistability and antiresonance
in integrator neurons was also studied recently \cite{Guantes2005}.

In two previous papers on the HH model driven by periodic
sequence of short stimuli, we have shown
that there is a transition between odd-only and all modes
at high frequencies \cite{Borkowski2009,Borkowski2010b}. This transition
is located between the locked-in regions $\textrm{2:1}$
and $\textrm{3:1}$. The notation $p:q$ means $q$ output spikes for every
$p$ input current pulses. The edges of individual modes
scale logarithmically in the vicinity of this singularity.
This theoretical analysis agrees well with experimental data \cite{Takahashi1990}.
A natural question to ask is whether we can identify
analogous parity transition, involving even-only modes
on one side and all modes on the other side of the transition.
A preliminary study \cite{Borkowski2009BMC}
showed that the even and odd modes
compete also near the main resonance.
In the following we analyze the resonance regime in detail,
looking for signatures of the even-all transition.

Our aim is to map the response diagram
of the HH neuron to a train of brief current pulses,
rather than emulate typical
\textit{in vivo} scenarios.
Stimuli in the form of brief pulses
are better suited to reveal the internal dynamics
of the neuron than signals varying on the time scale
of the main resonance.
In the case of the often used constant or sinusoidal input currents
the neuron dynamics is obscured by the drive
that is always different from zero.
Gaussian noise is added to the model in order to reveal
additional features of the model's dynamics.
Although this particular type of randomness
may not necessarily occur
in a neuronal system we believe similar results
will be obtained for any fast-varying irregular
component of the stimulus.

\section{The model and results}
We consider the stochastic HH model with the classic parameter set
and rate constants \cite{HH1952},
\begin{equation}
\label{HHnoise}
C\frac{dV}{dt}=-I_{Na}-I_K-I_L+I_{app}+C \xi(t),
\end{equation}
where noise is given by the Gaussian process
$<\xi_i(t)>=0$, 
$<\xi(t)\xi(t^\prime)> = 2D\delta(t-t^\prime)$,
and $D$ is expressed in $\textrm{mV}^2/\textrm{ms}$.
$I_{Na}$, $I_K$, $I_L$, and $I_{app}$,
are the sodium, potassium,
leak, and external current, respectively.
$C=1\mu\textrm{F/cm}^2$ is the membrane capacitance.
The input current is a periodic set
of rectangular steps of height $I_0$ and width $0.6 \textrm{ms}$,
which is an order of magnitude below the resonant
pulse width \cite{Borkowski2010} and does not interfere
with the neuron's internal dynamics.
The deterministic (stochastic) HH equations are integrated
using the fourth-order Runge-Kutta algorithm (the second-order stochastic
Runge-Kutta algorithm \cite{Honeycutt1992a}), respectively.
The simulations are carried out with the time step of $0.001 \textrm{ms}$
and are run for $400 \textrm{s}$, discarding the initial $4 \textrm{s}$.

\begin{figure}[t]
\includegraphics[width=0.46\textwidth]{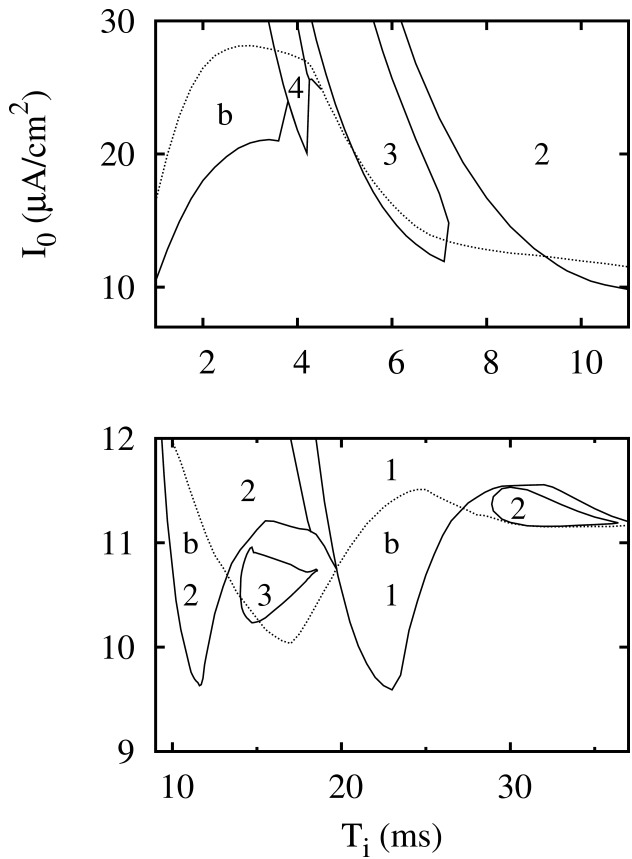}
\caption{Response diagram in the noiseless case
for stimulation by rectangular current
pulses of width $\tau=0.6 \textrm{ms}$ and height $I_0$.
The top (bottom) panel shows the high (intermediate) frequency
regime.
Bistable regions are marked "b". Some smaller bistable areas
remain unlabeled. They are limited by a dotted line from
above and a solid line from below.
Labels 1,2,3, and 4 mark
regions with $\textrm{1:1,2:1,3:1}$, and $\textrm{4:1}$
mode locking, respectively.
The resonance area near $T_i=17 \textrm{ms}$
is dominated by the $\textrm{3:1}$ state and higher order states.
The second resonance near $T_i=34 \textrm{ms}$
is dominated by the $\textrm{2:1}$ mode. Higher modes
appear near the threshold at $T_i \simeq 34\textrm{ms}$.
}
\label{fig1}
\end{figure}

Figure \ref{fig1} shows the response diagram
in the $T_i-I_0$ plane without noise, where $T_i$
is the stimulus period.
The main resonance is located at $T_i \simeq 17 \textrm{ms}$.
The second order resonance is present at $T_i\simeq 34 \textrm{ms}$
and the third one at $T_i \simeq 51 \textrm{ms}$ (not shown).
The dotted line separates the region with a single solution
from the area where two solutions coexist.
Bistable solutions appear between the resonant regions
near $T_i=11 \textrm{ms}$ and $T_i=23 \textrm{ms}$.
In these regimes the transition to excitability occurs
via the subcritical Hopf bifurcation.
The picture is more complicated
at the frequencies above 250 Hz, where
the quiescent state often coexists not with a limit cycle
but with a set of irregular orbits.  
The boundaries of bistable regions are determined
with a simple continuation algorithm.
The initial conditions of each run
with a new value of a bifurcation parameter
are equal to the end values
from the previous iteration.

The firing rate $f_o/f_i$
depends continuously on $I_0$
between the tip of the resonance at $T_i \simeq 17\textrm{ms}$ 
and the bistable area, which begins at $T_i \simeq 20 \textrm{ms}$.
Here $f_o$, $f_i$ is output, input frequency, respectively.
For $14 \textrm {ms} < T_i < 17 \textrm{ms}$ bistable intervals with
integer $p/q$ ratio alternate with irregular
response, whose long-time average is a continuous function
of $I_0$ and $T_i$.
Earlier, Clay reported \cite{Clay2003}
an irregular graded response near the excitation edge
within a revised HH model \cite{Clay1998}
stimulated by 1-ms rectangular current pulses
and trains of half-sine waves.
This variability was attributed to the deterministic
nonlinear dynamics of the model.

\begin{figure}[th]
\includegraphics[width=0.46\textwidth]{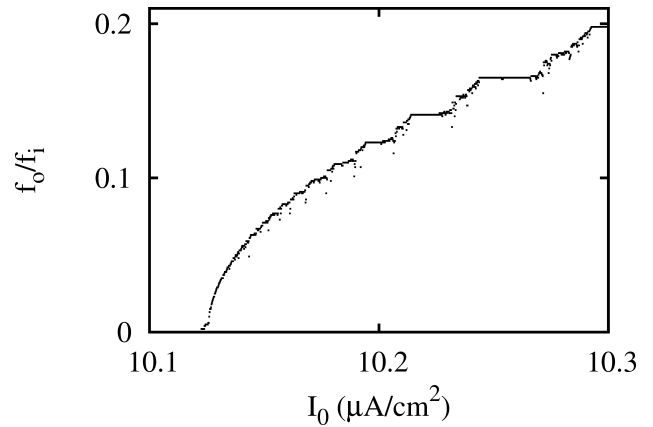}
\caption{The firing rate as a function
of the current pulse height at the resonance $T_i=17.5 \textrm{ms}$.
Here $\tau=0.6 \textrm{ms}$. Near the firing threshold
$f_o$ is approximately a square root function
of the pulse amplitude.
Further away from the threshold the dependence of $f_o$ on $I_0$
between the mode-locked states is irregular and nonmonotonic.
}
\label{fig2}
\end{figure}

\begin{figure}[th]
\includegraphics[width=0.46\textwidth]{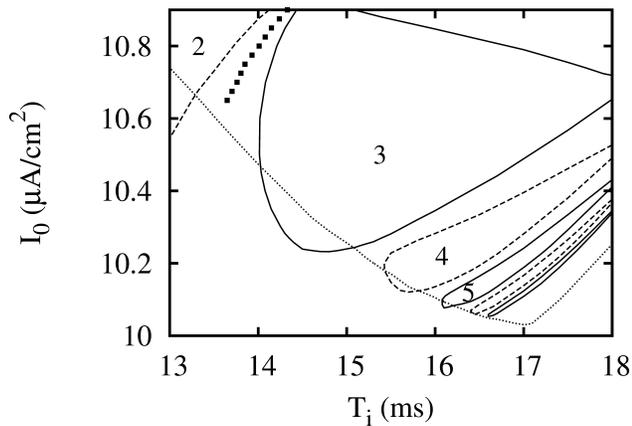}
\caption{Details of the response diagram at the resonance.
Borders of even (odd) mode-locked states are shown with
broken (continuous) lines respectively.
The dotted lines mark the transitions to repetitive firing.
Areas enclosed by the dotted line and continuous or broken lines
are bistable.
Numbers $\textrm{2,...,5}$ indicate the states $\textrm{2:1,...,5:1}$.
The states beyond $\textrm{7:1}$ are not shown.
Filled squares indicate the parity multimodal transition
between even-only and all modes.
}
\label{fig3}
\end{figure}

For $T_i \ge 17 \textrm{ms}$ the firing rate is approximately
a square root function of the deviation from the threshold current
amplitude (see Fig. \ref{fig2}).
This dependence is characteristic
of a saddle-node bifurcation \cite{Strogatz1994}.
The scaling $f_o/f_i \sim (I_0-I_th)^\beta$,
with $\beta=1/2$ is reminiscent of a mean-field second-order
phase transition, with $f_o/f_i$ playing the role of an order
parameter. The same exponent is obtained for a relaxation time
near $I_c\simeq 6.264\mu\textrm{A/cm}^2$
for a constant current $I$ \cite{Roa2007},
where $I_c$ is the value of $I$ at the saddle-node bifurcation.
Below $I_c$ the neuron returns to quiescence after
emitting a series of spikes. The relaxation time,
defined as the time from the first to the last spike,
diverges as $(I_c-I)^\Delta$, where $\Delta\simeq 1/2$.
Similar values of $\Delta$ were obtained
for the Morris-Lecar and FitzHugh-Nagumo models \cite{Roa2007}.

\begin{figure}[th]
\includegraphics[width=0.46\textwidth]{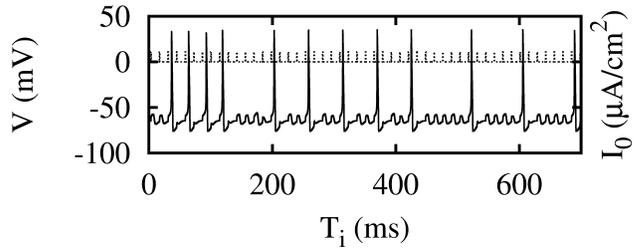}
\caption{Sample voltage trace in the chaotic regime near the transition between
even-only and all modes. The applied current pulses are shown with
dotted lines. There is only one odd multiple (7:1) of the input period
in this sample. Here $T_i=13.9 \textrm{ms}$ and $I_0=10.75 \mu\textrm{A/cm}^2$.
}
\label{fig4}
\end{figure}

Figure \ref{fig3} shows the bifurcation diagram near the tip
of the resonance. The synchronized states alternate
with irregular firing. There are small bistable regions
where the quiescent state coexists with a limit cycle.
We have calculated their boundaries for states up to order 5:1.
It is likely that they extend all the way to the tip
of the resonance.
The behavior in the intermediate regions is chaotic
due to a competition
between odd and even modes \cite{Borkowski2009}.
An example of $V(t)$ dependence for a point located
between states \textrm{2:1} and \textrm{3:1}
is shown in Fig. \ref{fig4}.
This sample contains only one odd multiple of the input period.
Odd interspike intervals vanish 
in the vicinity of the 2:1 state. Careful analysis
of the interspike interval histograms (ISIH) reveals
a transition between the set of even and the set of all modes.
This parity transition of ISIHs in the HH model
could be tested experimentally in a squid axon experiment,
similar to the odd-all transition
discovered earlier \cite{Takahashi1990,Borkowski2010b}.

\begin{figure}[th]
\includegraphics[width=0.46\textwidth]{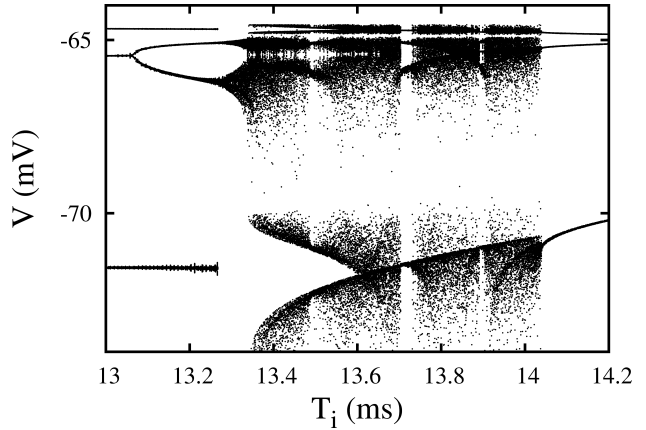}
\caption{The Poincare section of the membrane potential
near the bistable region and irregular regime
for current amplitude $I_0=10.65 \mu\textrm{A/cm}^2$.
The values of $V(t)$ in each run are recorded
at the points $t=nT_i$, where $n=1,2,...$.
The lowest and the highest flat branches on the left
belong to the $\textrm{2:1}$ states.
The middle one belongs to the steady state. It undergoes
a period doubling bifurcation on approach to the irregular
regime above $T_i=13.3\textrm{ms}$.
}
\label{fig5}
\end{figure}

The transition between quiescence and chaotic firing
marked by the dotted line
in the intermediate zones between the locked-in states
occurs via period doubling (see Fig. \ref{fig5}).
The lowest and the highest values of $V$ below
$T_i\simeq 13.24 \textrm{ms}$ belong to the $\textrm{2:1}$ limit cycle.
The line slightly below $V=-65 \textrm{mV}$, splitting above
$T_i=13 \textrm{ms}$, is associated with the steady state.

\begin{figure*}[th]
\centering
\includegraphics[width=0.46\textwidth]{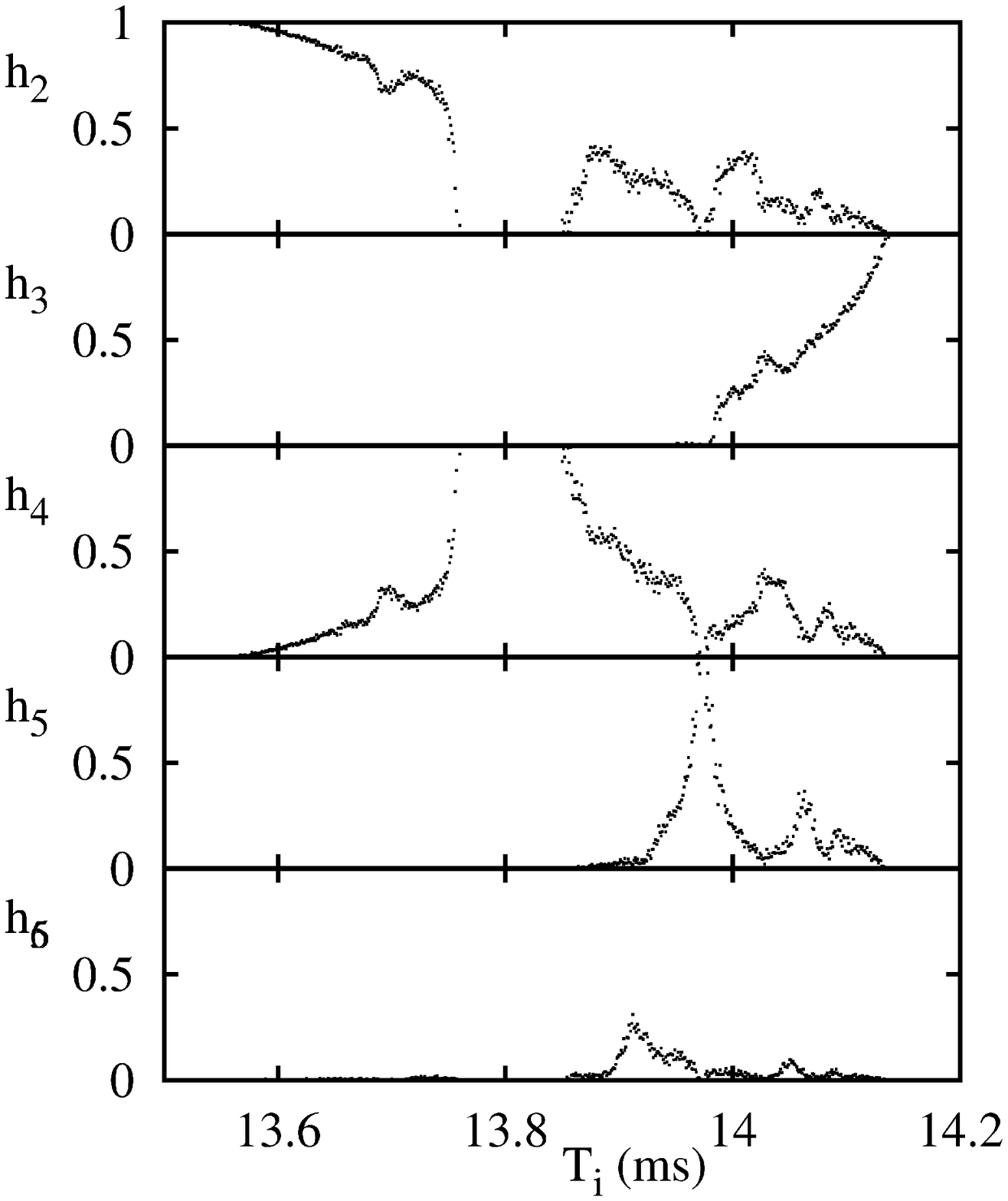}
\hskip 1cm
\includegraphics[width=0.46\textwidth]{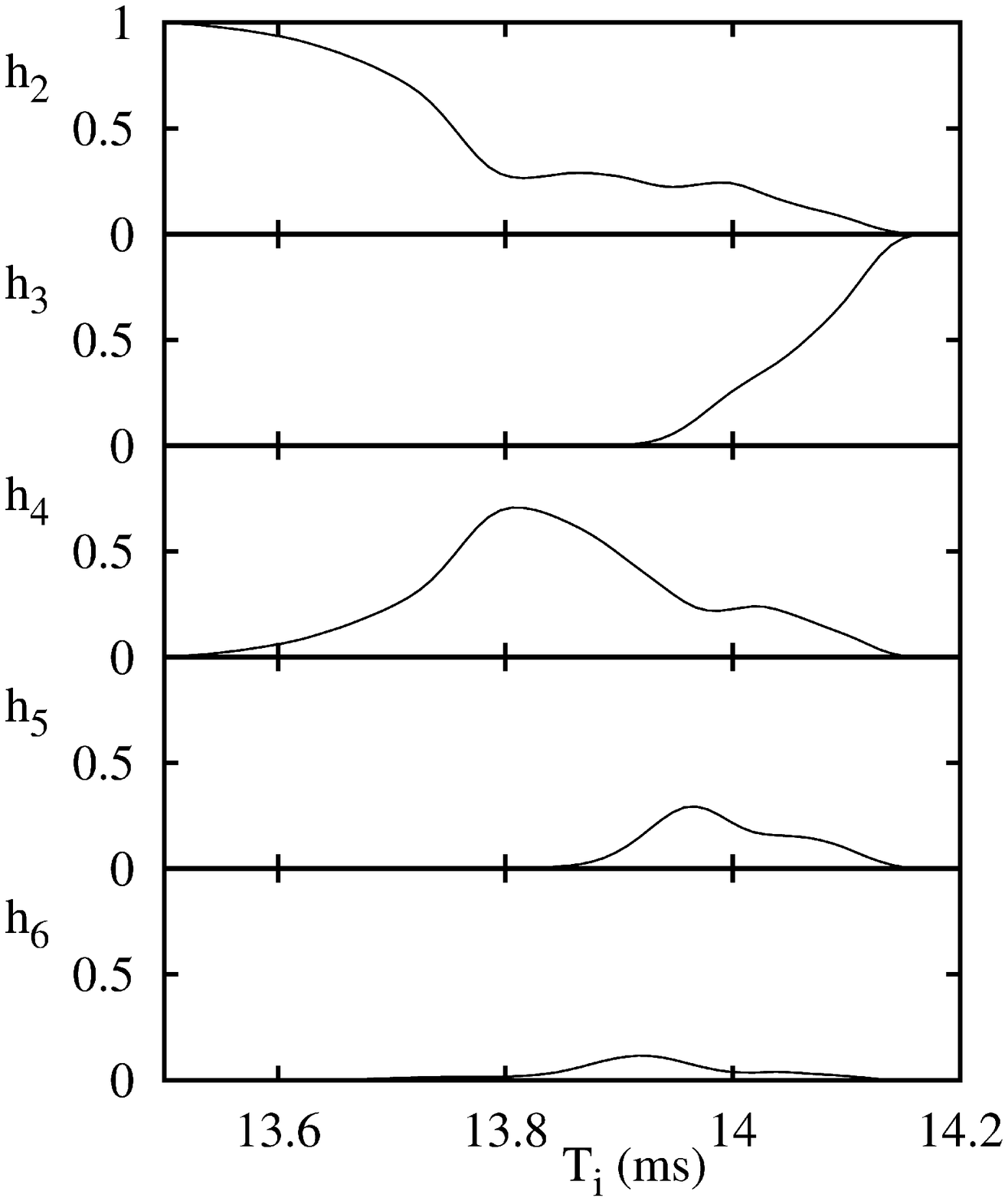}
\caption{Histograms of the lowest order modes for $D=0$ (left)
and $D=10^{-5}$ (right) at $I_0=10.75\mu\textrm{A/cm}^2$.
This is an irregular firing regime between
the states $\textrm{2:1}$ and $\textrm{3:1}$.
Data in the noisy case are Bezier-smoothed averages.
}
\label{fig6}
\end{figure*}

The even, odd multiples of $T_i$ dominate near the states with even, odd
$p/q$ ratio, respectively (see Fig. \ref{fig6}, left). This is a general
property of regimes of irregular firing.
Starting from the $\textrm{2:1}$ mode, the histogram weight is gradually transferred
to the $\textrm{4:1}$ mode, which dominates for $T_i$
slightly below the even-all transition point $T_{ea}$.
On the other side of $T_{ea}$, $h_5=1$ over a narrow interval.

The histogram of the dominant modes for $D>0$ is shown in Fig. \ref{fig6} (right).
Some weight is now transferred to lower modes.
The $\textrm{4:1}$ and $\textrm{5:1}$ modes are still well pronounced over a range of $T_i$
but they are now mixed with $\textrm{2:1}$ and $\textrm{3:1}$ modes, respectively.
Noise extends the range of presence of high order modes
around $T_{ea}$, as expected.

\begin{figure}[th]
\includegraphics[width=0.46\textwidth]{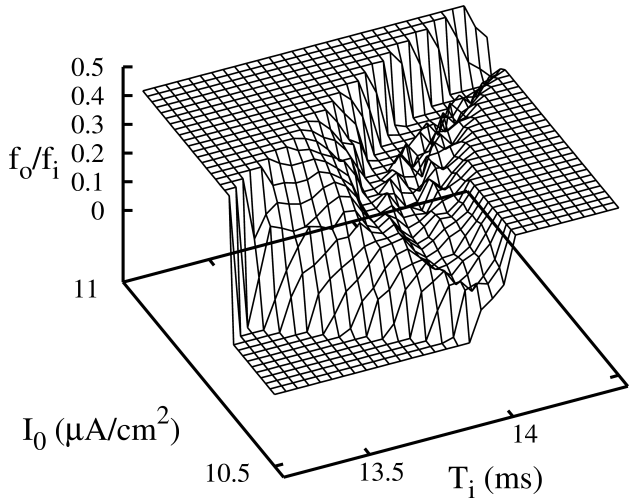}
\caption{Minimum of the firing rate between the $\textrm{2:1}$
and $\textrm{3:1}$ states for the noiseless stimulus.
The state $\textrm{4:1}$ is located near the $\textrm{2:1}$ mode.
}
\label{fig7}
\end{figure}

The average firing rate between nearby $\textit{n}\textrm{:1}$
and $(\textit{n}\textrm{+1):1}$ states
often has a narrow local minimum due
to the presence of slower modes. The minima are more pronounced near
the excitation threshold along the left edge
of the resonance tip in Fig. \ref{fig3},
where $T_i < 17 \textrm{ms}$. One such valley
is shown in Fig. \ref{fig7}. Here the $\textrm{4:1}$ mode
dominates in a narrow range of parameter values
below the parity transition.

\begin{figure}[th]
\includegraphics[width=0.46\textwidth]{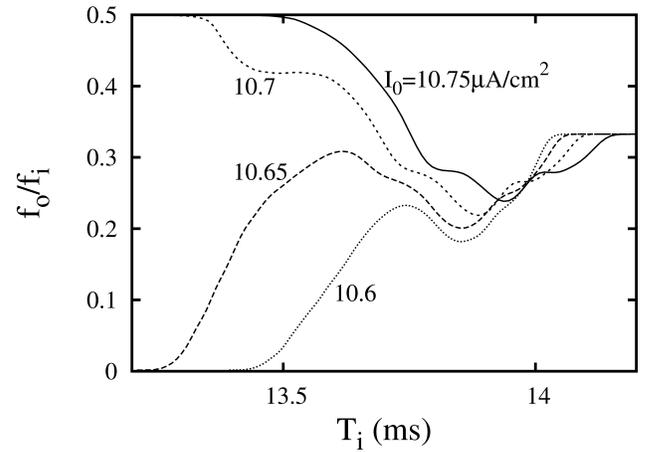}
\caption{The firing rate as a function of $T_i$ in the intermediate region
between the $\textrm{2:1}$ and $\textrm{3:1}$ states for $D=10^{-5}$
and $T_i=10.75\textrm{ms}$. The competition of odd and even modes
manifests itself as a minimum of the firing rate.
}
\label{fig8}
\end{figure}

The firing rate minimum is robust to small levels of noise, as can be seen
in Fig. \ref{fig8}. 
The minimum of $f_o$ occurs for $T_i\gtrsim T_{ea}$,
where modes of both parity are available.
In the odd-all transition
of the high-frequency regime \cite{Borkowski2009}
the minimum $f_o$ was similarly located close to the transition,
on the side of all modes.

\begin{figure}[th]
\includegraphics[width=0.47\textwidth]{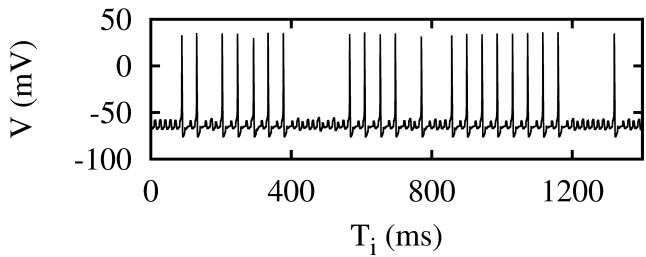}
\caption{Sample voltage trace in the bistable regime
of the state $\textrm{3:1}$
under the influence of small noise. The most common form of response
consists of 3:1 bursts separated by longer silent intervals.
Here $T_i=13.9 \textrm{ms}$, $I_0=10.75 \mu\textrm{A/cm}^2$, and $D=10^{-3}$.
}
\label{fig9}
\end{figure}

It is well known that the HH neuron
has a tendency to spike in bursts when subjected
to a noisy stimulus in a bistable regime.
If the deterministic system is prepared
in one of high order bistable states from Fig. \ref{fig3},
the addition of noise results in slow bursts,
where the ISI within a burst is a high integer multiple of $T_i$.
An example is shown in Fig. \ref{fig9},
where each burst's ISI is equal to $3T_i$.
We expect slow bursts of order 6:1 and higher to be found
in a more detailed calculation. 

\begin{figure}[th]
\includegraphics[width=0.4\textwidth]{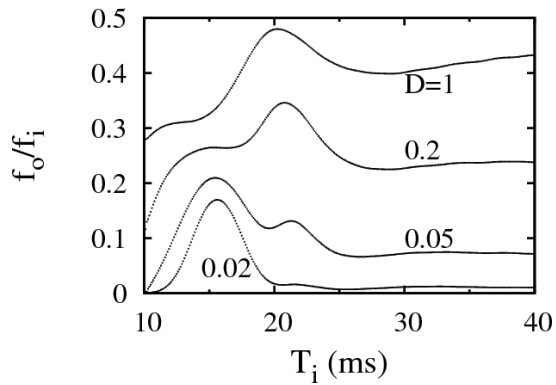}
\caption{The firing rate as a function of $T_i$
for different noise levels and $I=10 \mu\textrm{A/cm}^2$,
about $0.1 \mu\textrm{A/cm}^2$ below the deterministic threshold.
}
\label{fig10}
\end{figure}

For stimuli below the deterministic threshold
noise activates firing with maximum response
at the resonant frequency (see Fig. \ref{fig10}).
When noise is larger, excitations into the
$\textrm{2:1}$ state occur more frequently and the maximum
of the firing rate shifts toward $T_i=20 \textrm{ms}$.
This maximum is a result of the interplay
between the $\textrm{1:1}$ mode and the $\textrm{2:1}$ mode.
The relative frequency of participation
of the $\textrm{2:1}$ mode grows sharply
for intermediate noise levels in the neighborhood
of the bistable regime for $T_i \simeq 20 \textrm{ms}$.

For signals with a periodic subthreshold component noise
may play the role of a frequency selector.
In a network with some elements firing
in unison at the resonant frequency the remaining
uncorrelated neurons also play an important role.
The intensity of their background activity
may select the firing rate of a network.

\begin{figure}[th]
\includegraphics[width=0.4\textwidth]{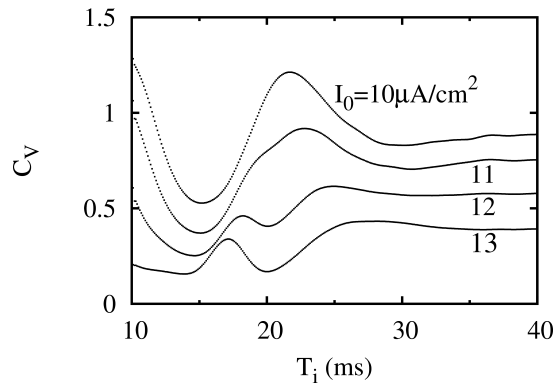}
\caption{The coefficient of variation as a function of $T_i$
for different pulse amplitudes. The noise intensity
if fixed at $D=0.2$.
The local minimum at larger $I_0$ follows the left edge
of the $\textrm{1:1}$ regime.
}
\label{fig11}
\end{figure}

The dependence of $C_V$ on $T_i$ for fixed noise
intensity is shown in Fig. \ref{fig11}.
As expected, the most coherent response occurs
near the resonance.
The minimum near $T_i=20 \textrm{ms}$ for suprathreshold
inputs follows the left edge of the $\textrm{1:1}$ region
from Fig. \ref{fig1}.
For stimuli below threshold there is a deep minimum
at the resonance and a large maximum in the bistable
regime for $T_i > 20 \textrm{ms}$.

For suprathreshold stimulation in the resonance regime the firing rate
in general increases monotonically as a function of $D$ with the exception
of states having $p/q$ values close to 1 (see Fig. \ref{fig11}).
Starting in the $\textrm{1:1}$ state at $D=0$,
increasing noise slows down the response
by annihilating  some of the action potential spikes.
The minimum $f_o/f_i$ is reached at an intermediate $D$.
In the limit of large noise $f_o$ approaches the inverse
of the refractory period.
The $C_V$ maxima are caused by redistribution
of histogram weight among several principal modes.
This effect was described earlier as stochastic
coherence antiresonance \cite{Borkowski2010b}.
Maximization of spike train incoherence was shown
earlier to occur in the FitzHugh-Nagumo \cite{Lacasta2002}
and leaky integrate-and-fire \cite{Lindner2002} models.
However in those studies the maxima of $C_V$ were not
accompanied by the minima of $f_o/f_i$.

\begin{figure}[th]
\includegraphics[width=0.4\textwidth]{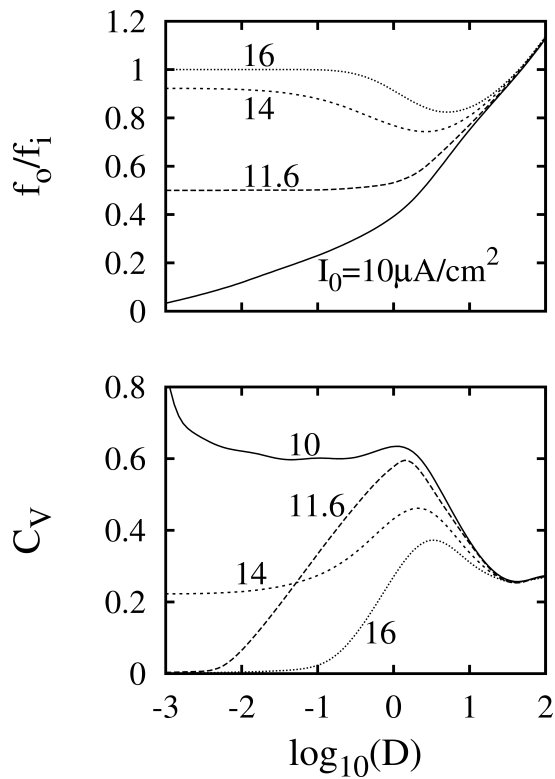}
\caption{The firing rate and $C_V$ as a function of $D$
for several values of $I_0$ and $T_i=17 \textrm{ms}$.
Maximum of $C_V$ at intermediate values of $D$ is due
to the multimodal distribution of interspike intervals.
}
\label{fig12}
\end{figure}

\begin{figure}[ht]
\includegraphics[width=0.48\textwidth]{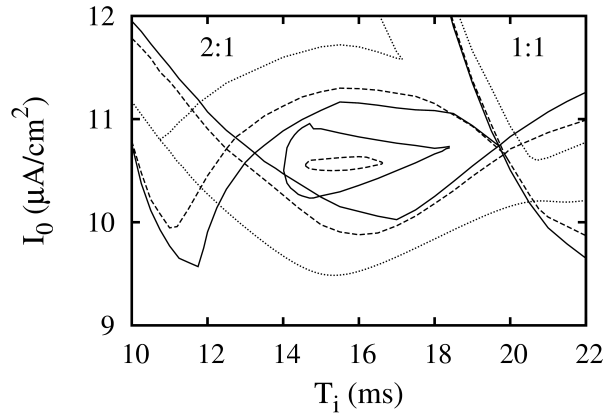}
\caption{The response diagram in the presence of noise.
Solid lines are the boundaries of the main mode-locked
states without noise (see also Fig. \ref{fig1}).
The broken and dotted lines are boundaries for
$D=10^{-3}$ and $10^{-2}$, respectively.
}
\label{fig13}
\end{figure}

\begin{figure}[ht]
\includegraphics[width=0.48\textwidth]{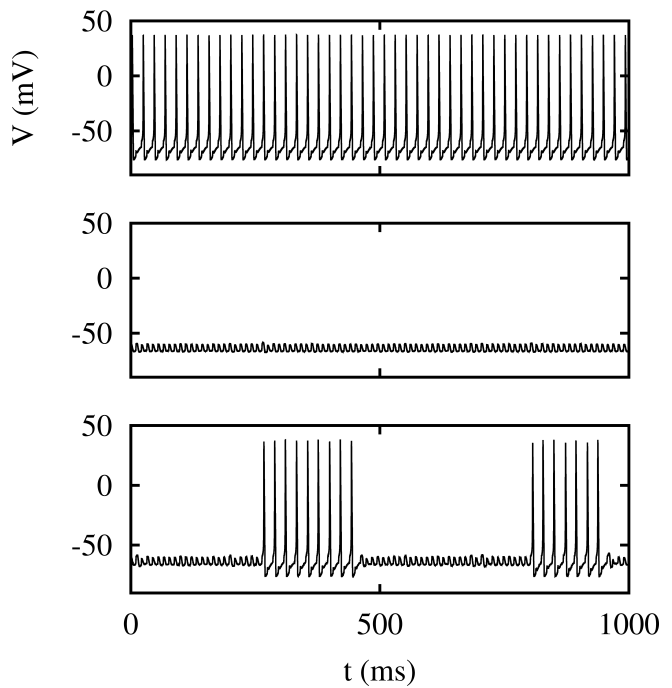}
\caption{Sample dependence of $V(t)$ for three intensities of noise,
from top to bottom: $D=0.001, 0.01$, and $0.03$. The spiking action
is switched off for intermediate values of $D$.
Here $T_i=11\textrm{ms}$, $I_0=10.5\mu\textrm{A/cm}^2$.
}
\label{fig14}
\end{figure}

Figure \ref{fig13} shows the response diagram
in the presence of noise.
The tip of the resonance is broadened and shifted
to higher frequencies.
The $\textrm{3:1}$ state vanishes completely
for $D$ slightly larger than $10^{-3}$. Higher-order
states are more sensitive to noise and are quickly washed out.
The bistable area at the edge of the \textrm{2:1} state
shrinks more gradually. When bistability
is finally eliminated, $f_o$ remains discontinuous
over part of that border.
For $D=10^{-2}$ the discontinuity occurs below $T_i\simeq 10.5\textrm{ms}$.
Above $T_i\simeq 10.5\textrm{ms}$ the firing rate is
a continuous function of $I_0$.
The behavior in the immediate vicinity of $T_i=10.5\textrm{ms}$
seems to be weakly irregular. A more detailed study would be
needed for a proper description of this area.
The loss of stability by the \textrm{2:1} state
in favor of the quiescent state and further transition
to bursting are illustrated in Fig. \ref{fig14}.

\begin{figure}[ht]
\includegraphics[width=0.48\textwidth]{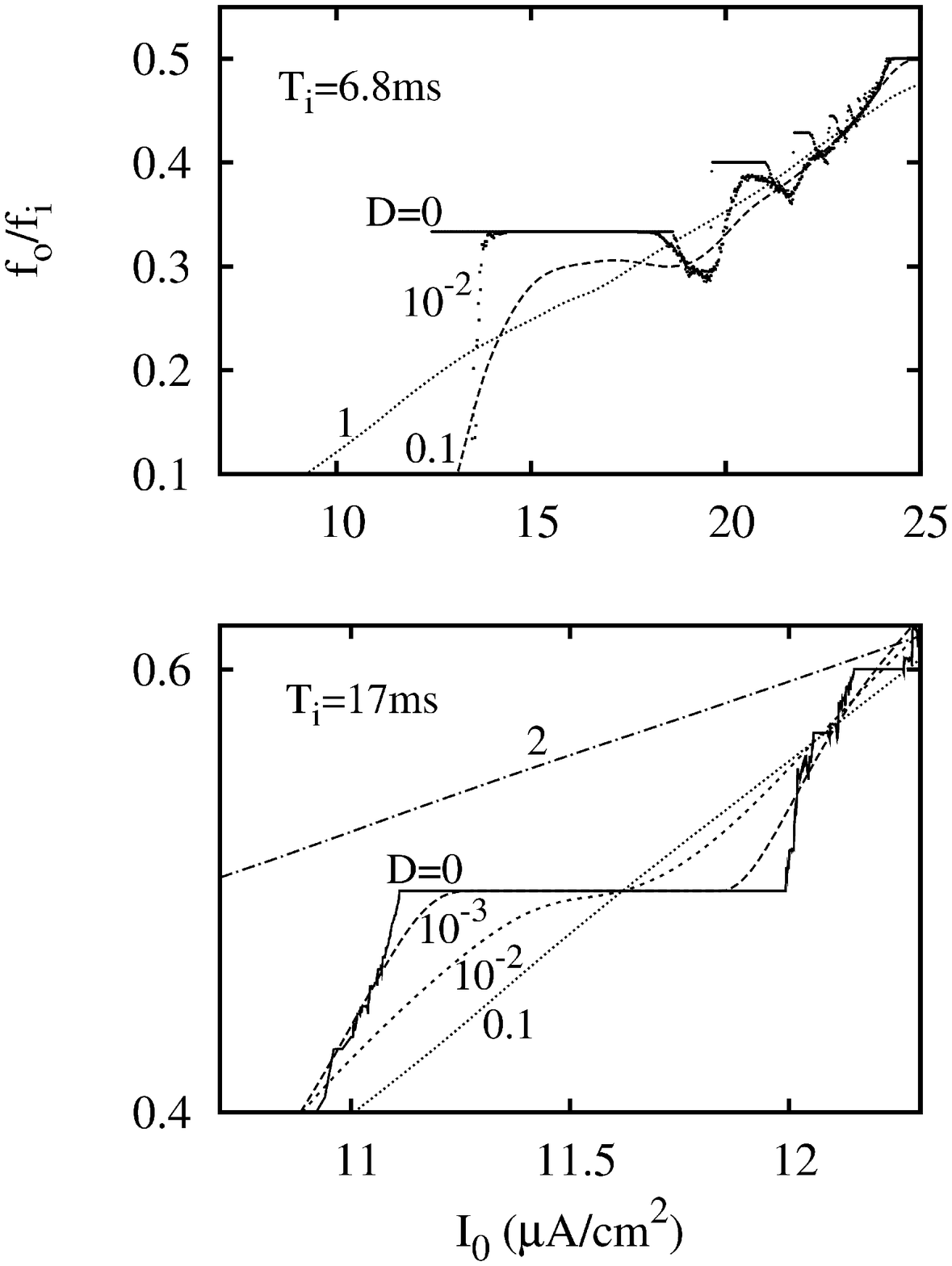}
\caption{Typical behavior
of the average firing rate in the regime of high (top)
and moderate frequency (bottom) for different
noise levels.
At high frequencies and small noise
the response is initially slowed down
over an entire plateau, starting from both edges.
At the lower diagram $f_o$ is held steady in the central
part of the plateau, decreasing at the left edge
and increasing at the right edge.
The average firing rate in the central part of the plateau
is preserved up to $D\simeq 0.1$.
The irregular nonmonotonic behavior in the deterministic case
is smoothed out for very small levels of noise.
}
\label{fig15}
\end{figure}

The reaction to noise at high frequencies
is qualitatively different from behavior
at near-resonant frequencies (see Fig. \ref{fig15}).
In the top panel of Fig. \ref{fig15}
the firing rate drops almost everywhere for small $D$.
This is easy to understand if we recall
that the odd-all multimodal transition \cite{Borkowski2009,Borkowski2010b}
occurs just below $I_0=20\mu\textrm{A/cm}^2$.
The slow modes are easily available in this regime
and small perturbations suffice to switch
the system among different trajectories.
The firing rate drops over the entire \textrm{3:1} plateau
due to the appearance of the \textrm{5:1}, \textrm{7:1}
and other odd modes. For larger $D$ also the even modes
are sampled by the system and $f_o/f_i$ increases.

At moderate, near-resonance frequencies,
the central part of the \textrm{2:1} plateau
in the lower part of Fig. \ref{fig15} is more robust to noise.
The average $f_o$ is preserved for $D$ up to $0.1$.
Noise induces both the \textrm{1:1} mode as well as
the $\textrm{3:1}$ and slower modes.
However the participation rate of the $\textrm{1:1}$ mode is balanced
by the slower modes in such a way that average $f_o/f_i$ remains close to 2.
This type of resilience to noise, characteristic
of the resonance regime, can also be seen
in other mode-locked states.

\begin{figure}[th]
\includegraphics[width=0.48\textwidth]{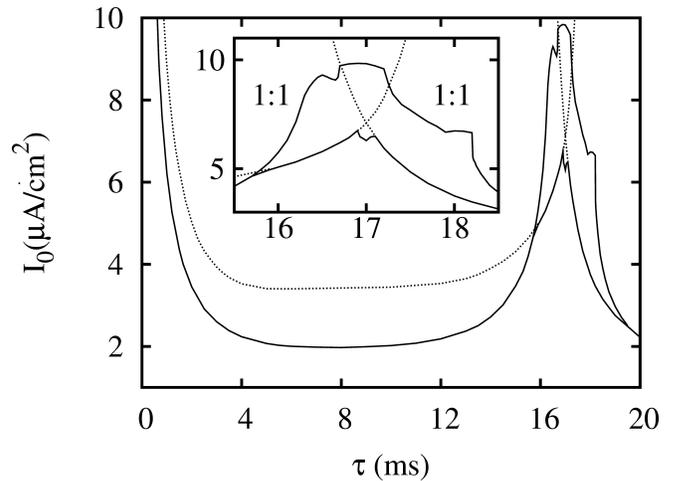}
\caption{
Bifurcation diagram near the excitation threshold
as a function of the current pulse width.
The input period is set at $T_i=17\textrm{ms}$.
At $\tau=17\textrm{ms}$ the input current is
constant. The details of the antiresonant limit
with bistable behavior are shown in the inset.
Solid lines mark the excitation threshold.
The dotted lines are the boundaries of the $\textrm{1:1}$ state.
}
\label{fig16}
\end{figure}

Finally, let us briefly describe the role played by the current
pulse width $\tau$.
Figure \ref{fig16} shows the excitation threshold
as a function of $\tau$ for a resonant drive, $T_i=17\textrm{ms}$.
Initially the threshold decreases as the inverse of $\tau$.
For $\tau > 8 \textrm{ms}$ the tip of the resonance
(shown in Fig. \ref{fig3})
gradually becomes narrower and vanishes.
The threshold rises again and reaches
maximum near $\tau=17\textrm{ms}$, when $I(t)=\textrm{const}$.
Bistability appears above $\tau\simeq 16\textrm{ms}$.
It is clear that stimulation by a constant current
forces the neuron into an antiresonant regime.
The bifurcation diagram in Fig. \ref{fig16}
implies that the addition of any charge-unbalanced component
to the constant stimulus takes the system away
from this antiresonant limit.
The precise nature of such a component, whether deterministic
or stochastic, is not important. What matters is
the amount of charge delivered within approximately $4\textrm{ms}$
from the stimulus onset.

\section{Conclusions}
We studied the response of the HH neuron to a periodic pulse current
with a Gaussian noise. The global bifurcation diagram
in the $T_i-I_0$ plane has a rich structure
near the excitation threshold, where
resonant regimes alternate with antiresonant ones.
The model is bistable between the resonances and
in the limit of high-frequency stimulation.

The firing rate is a continuous function of $I_0$
for $T_{res} \lesssim T_i < 20\textrm{ms}$.
The scaling of the firing rate with $(I_0-I_{th})^{1/2}$
is a signature of a saddle-node bifurcation at the threshold.
For $T_i \lesssim T_{res}$ bistable regions
are separated by areas of irregular response
with no well defined threshold
and approximately continuous dependence
of $f_o/f_i$ on input parameters.
As $T_i$ approaches $T_{res}$ from below,
the subcritical Hopf bifurcation gradually softens.
Bistable regions occupy smaller portions
of the parameter space and disappear before $T_i=T_{res}$.
The change of the type of neuronal excitability is
important in the context of preventing the so called dynamical diseases,
such as epilepsy, Alzheimer's, and Parkinson's disease.
It was shown in the HH \cite{Wang2007,Xie2008a}
and the Hindmarsh-Rose \cite{Xie2008} models
that this could be obtained by introducing an additional
control function.
We demonstrated that the threshold behavior
can also be altered by selecting the frequency
and amplitude of external current.

For subthreshold stimuli noise enables spiking
in the vicinity of the resonance.
The stimulus frequency, for which the maximum firing
rate is obtained, depends on the magnitude of noise.
For large $D$, the maximum $f_o$ occurs for $T_i>T_{res}$.
It would be interesting to investigate the same regime
in a HH network where the connectivity pattern of neurons
as well as their individual properties
might lead to the emergence of subpopulations of neurons
firing with different average frequencies in response
to a correlated input with background noise.
It was shown earlier by several authors that both
the background noisy activity \cite{Ho2000,Chance2002}
and correlated inputs \cite{Stevens1998} are important
in explaining the neuronal response \textit{in vivo}.
Qualitative results of this work do not depend on the precise functional form
of the current pulse provided the width of each pulse
does not exceed 4 ms.
This invariance of the bifurcation diagram in the $T_i-I_0$ plane
is not surprising because for short pulses the HH neuron's threshold
is determined by the amount of charge delivered per pulse \cite{Koch1999}.

We have also found a new even-all multimodal transition
occurring between the states 2:1 and 3:1 close
to the main resonance.
For input period $T_i$ below this singularity
only even response modes exist.
Both even and odd modes appear above the transition.
This effect is accompanied by a minimum of the firing
rate, located close to the parity transition, on the side
of all modes. Similar transitions may exist in other excitable
systems.

Our results are also relevant to studies of auditory
nerve fiber responses to electric
stimulation \cite{Matsuoka2000,Macherey2007,OGorman2009,OGorman2010}.
At low stimulation rates auditory nerve fibers
fire regularly and are locked in to
applied stimulus. At high stimulation rates
these fibers respond irregularly.
Some researchers attributed this effect to physiological
noise \cite{Morse1996,Moss1996}.
However, an analysis within the FitzHugh-Nagumo model
showed that the firing irregularities
at high frequencies may be caused by deterministic dynamical
instability \cite{OGorman2009,OGorman2010}.
We have found similar instabilities
in the HH model. They appear at high stimulation frequencies
and along the excitation threshold. Both of these regimes
are relevant to studies of hearing sensitivity.

\acknowledgments
Computations were performed in the Computer Center
of the Tri-city Academic Computer Network in Gdansk.

\bibliography{Borkowski2010c}
\end{document}